\documentclass[12pt]{article}
\usepackage{graphics}
\usepackage{bm}
\usepackage{graphicx}
\usepackage{amsmath}
\usepackage{amssymb}
\usepackage{amstext}
\usepackage{epstopdf}
\usepackage{amscd}
\usepackage{amsfonts}
\usepackage{appendix}
\usepackage{color}
\usepackage{wasysym}
\usepackage{textcomp}
\DeclareMathAlphabet{\EuFrak}{U}{euf}{m}{n}
\DeclareMathAlphabet{\EuScript}{U}{eus}{m}{n}

\newcommand{\nd}{\noindent}

\newcommand{\be}{\begin{equation}}
\newcommand{\ee}{\end{equation}}
\newcommand{\ben}{\begin{eqnarray}}
\newcommand{\een}{\end{eqnarray}}

\title{{{\bf Dimensionally regularized Boltzmann-Gibbs Statistical Mechanics 
and two-body Newton's gravitation}}}
\author{{\small{D. J. Zamora$^{1,4}$, M. C. Rocca$^{1,2,4}$,A. Plastino$^{1,4,5}$,
G. L. Ferri $^3$}}, \\
\small{$^1$ Departamento de F\'{\i}sica,
Universidad Nacional de La Plata,}\\
\small{$^2$ Departamento de Matem\'{a}tica,
Universidad Nacional de La Plata,}\\
\small{$^3$Fac. de C. Exactas-National University La Pampa,} \\
\small{Peru y Uruguay, Santa Rosa, La Pampa, Argentina}\\
\small{$^4$ Consejo Nacional de Investigaciones Cient\'{\i}ficas
y Tecnol\'{o}gicas}\\
\small{(IFLP-CCT-CONICET)-C. C. 727, 1900 La Plata -
Argentina}\\\small{$^5$  SThAR - EPFL, Lausanne, Switzerland}}

\date{\today}

\begin{document}

\maketitle

\begin{abstract}

It is believed that the canonical gravitational 
partition function $Z$ associated to the classical Boltzmann-Gibbs (BG) distribution 
$\frac {e^{-\beta H}} {{\cal Z}}$ cannot be constructed because the integral
needed for building up $Z$ includes an exponential and thus diverges at the origin. 
 We show here that, by recourse to  1) the analytical extension treatment 
obtained for the first time ever,
by Gradshteyn and Rizhik, via an 
appropriate formula for such case
and 2) the dimensional regularization approach of Bollini and Giambiagi's (DR),
one can indeed  obtain  finite gravitational results employing the BG distribution.
The BG treatment is considerably more involved than
its Tsallis counterpart. The latter needs only dimensional regularization,
the former requires, in addition, analytical extension.
\nd PACS: 05.20.-y, 02.10.-v\\
\nd KEYWORDS: Boltzmann-Gibbs distribution, divergences,  
dimensional regularization, specific heat.

\end{abstract}

\renewcommand{\theequation}{\arabic{section}.\arabic{equation}}

\section{Introduction}

DR \cite{tq1,tp1} constitutes one of the
greatest advances in the theoretical physics of the
last 45 years, with applications in several branches
of physics (see, for instance, \cite{dr1}-\cite{dr54}. \vskip 2mm

\nd It is commonly believed that the classical Boltzmann-Gibbs (BG) 
probability distribution can not yield finite results because the associated partition function  
${\cal Z}$ in $\nu$ dimensions diverges \cite{lb,grav1}, as one has ($m$ and $M$ are the masses involved,  $G$ the gravitation constant, $\beta$ the inverse temperature, and $x$-$p$ the phase-space coordinates) 

\begin{equation}
\label{newep3.1}
{\cal Z}_\nu=\int\limits_Me^{-\beta\left(\frac {p^2} {2m}-
\frac {GmM} {r}\right)}d^\nu xd^\nu p,
\end{equation}
with a positive exponential. However, such belief does not take into account the
possibility of  analytical extensions, that would take care of divergences, e.g., at the origin. 

\vskip 2mm  \nd It has been shown in Ref.\cite{z}, for first time ever, that
${\cal Z}$ can be calculated for Tsallis entropy using the
40-years old DR technique.

\vskip 2mm  \nd  Why are we insisting on this issue if it has been already solved?.
The issue needs revisiting because it does not work for
$q=1$, that is, for the Boltzmann-Gibbs statistics, due to the fact that we there face an 
exponential divergence.
In this paper we  report on how to  overcome this problem by judicious
use of an appropriate combination of DR plus analytical
extension. This produces the first ever ´BG partition function
for the two-body gravitational problem. We remark that
the N-body gravitational problem has not yet been solved
and constitutes a frontier research problem in Celestial Mechanics.

\vskip 2mm

\nd It is well known that, at a quantum field theory level,  DR can not cope with the gravitational field, since it is non-renormalizable. Our present challenge is quite different, though, because we deal with Newton's gravity at a {\it classical} level. 
  
\setcounter{equation}{0}

\section{Analytic extension}
In this section we collect a set of mathematical results that will be needed afterwards. This Section may be omitted at a first reading. 
We must now keep in mind that we are dealing with the integral of an 
exponentially increasing function given by (\ref{newep3.1}). We resort to 
Ref. \cite{gr}, and following it we consider a useful integral, that will greatly help with our inquires, after judicious specializations of it. This integral reads   
\begin{equation}
\label{ep2.1}
\int\limits_0^{\infty}x^{\nu-1}(x+\gamma)^{\mu-1}
e^{-\frac {\beta} {x}}dx=\beta^{\frac {\nu-1} {2}}
\gamma^{\frac {\nu-1} {2}+\mu}
\Gamma(1-\mu-\nu)
e^{\frac {\beta} {2\gamma}}
W_{\frac {\nu-1} {2}+\mu,-\frac {\nu} {2}}\left(
\frac {\beta} {\gamma}\right),
\end{equation}
$|\arg(\gamma)|<\pi$, $\Re(1-\mu-\nu)>0$, where 
 $W$ is {\it one} of the two  Whittaker functions. 
One does not require $\Re\beta>0$, as emphasized by Gradshteyn and Rizhik \cite{gr} (see 
figure in page 340, eq. (7), called ET II 234(13)a, where reference is made  to  \cite{gr1} (Caltech's 
Bateman Project).  The last letter ''a'' indicates that analytical extension has been performed. 
Choosing $\mu=1$ above we find
\begin{equation}
\label{ep2.2}
\int\limits_0^{\infty}x^{\nu-1}
e^{-\frac {\beta} {x}}dx=\beta^{\frac {\nu-1} {2}}
\gamma^{\frac {\nu+1} {2}}
\Gamma(-\nu)
e^{\frac {\beta} {2\gamma}}
W_{\frac {\nu+1} {2},-\frac {\nu} {2}}\left(
\frac {\beta} {\gamma}\right),
\end{equation}
valid for  $\nu\neq 0,-1,-2,-3,.....$
Additionally \cite{gr},
\begin{equation}
\label{ep2.3}
W_{\frac {\nu+1} {2},-\frac {\nu} {2}}\left(
\frac {\beta} {\gamma}\right)=
M_{\frac {\nu+1} {2},\frac {\nu} {2}}\left(
\frac {\beta} {\gamma}\right)=
\left(\frac {\beta} {\gamma}\right)^{\frac {\nu+1} {2}}
e^{-\frac {\beta} {2\gamma}},
\end{equation}
where  $M$ stands for the {\it other} Whittaker function.  Thus, 
 
\begin{equation}
\label{ep2.4}
\int\limits_0^{\infty}x^{\nu-1}
e^{-\frac {\beta} {x}}dx=\beta^{\nu}
\Gamma(-\nu)
\end{equation}
an integral that can be evaluated for all
 $\nu=1,2,3,....$ by recourse to the dimensional regularization technique \cite{tq1,tp1}. 
Changing now $\beta$ by $-\beta$ in (\ref{ep2.1})
 we have
\begin{equation}
\label{ep2.5}
\int\limits_0^{\infty}x^{\nu-1}(x+\gamma)^{\mu-1}
e^{\frac {\beta} {x}}dx=(-\beta)^{\frac {\nu-1} {2}}
\gamma^{\frac {\nu-1} {2}+\mu}
\Gamma(1-\mu-\nu)
e^{-\frac {\beta} {2\gamma}}
W_{\frac {\nu-1} {2}+\mu,-\frac {\nu} {2}}\left(-
\frac {\beta} {\gamma}\right).
\end{equation}
Once again we choose $\mu=1$ and have
\begin{equation}
\label{ep2.6}
\int\limits_0^{\infty}x^{\nu-1}
e^{\frac {\beta} {x}}dx=(-\beta)^{\frac {\nu-1} {2}}
\gamma^{\frac {\nu+1} {2}}
\Gamma(-\nu)
e^{-\frac {\beta} {2\gamma}}
W_{\frac {\nu+1} {2},-\frac {\nu} {2}}\left(-
\frac {\beta} {\gamma}\right),
\end{equation}
valid for $\nu\neq 0,-1,-2,-3,.....$ One now faces

\begin{equation}
\label{ep2.7}
W_{\frac {\nu+1} {2},-\frac {\nu} {2}}\left(-
\frac {\beta} {\gamma}\right)=
M_{\frac {\nu+1} {2},\frac {\nu} {2}}\left(-
\frac {\beta} {\gamma}\right)=
\left(-\frac {\beta} {\gamma}\right)^{\frac {\nu+1} {2}}
e^{\frac {\beta} {2\gamma}},
\end{equation}
and
\begin{equation}
\label{ep2.8}
\int\limits_0^{\infty}x^{\nu-1}
e^{\frac {\beta} {x}}dx=(-\beta)^{\nu}
\Gamma(-\nu)
\end{equation}
tantamount to changing  $\beta$ by $-\beta$ in (\ref{ep2.4}).
We have thus shown a rather interesting fact. Restriction 
of analytical extension (AE) of (\ref{ep2.1}) equals 
AE of the restriction of that relation. This reconfirms that 
 Gradshteyn and Rizhik's AE is indeed  correct.
Eq.  (\ref{ep2.8}) displays a cut at 
$\Re\beta>0$. One can then choose  
 $(-\beta)^\nu=e^{i\pi\nu}\beta^\nu$,
$(-\beta)^\nu=e^{-i\pi\nu}\beta^\nu$, or
$(-\beta)^\nu=\cos(\pi\nu)\beta^\nu$. We select the last possibility and obtain  
\begin{equation}
\label{ep2.9}
\int\limits_0^{\infty}x^{\nu-1}
e^{\frac {\beta} {x}}dx=\cos(\pi\nu)\beta^{\nu}
\Gamma(-\nu),
\end{equation}
an important result that we will use in Section 3. \vskip 3mm
\nd From  \cite{gr} we note that
\begin{equation}
\label{ep2.10}
\int\limits_0^{\infty}x^{\nu-1}
e^{-\beta x^2-\gamma x}dx=
(2\beta)^{-\frac {\nu} {2}}\Gamma(\nu)
e^{\frac {\gamma^2} {8\beta}}
D_{-\nu}\left(\frac {\gamma} {\sqrt{2\beta}}\right),
\end{equation}
where $D$ is the parabolic-cylinder function.
Selecting $\gamma=0$ one finds
\begin{equation}
\label{ep2.11}
\int\limits_0^{\infty}x^{\nu-1}
e^{-\beta x^2}dx=
(2\beta)^{-\frac {\nu} {2}}\Gamma(\nu)
D_{-\nu}(0).
\end{equation}
Since
\begin{equation}
\label{ep2.12}
D_{-\nu}(0)=\frac {2^{-\frac {\nu} {2}}\sqrt{\pi}}
{\Gamma\left(\frac {\nu+1} {2}\right)},
\end{equation}
we find

\begin{equation}
\label{ep2.13}
\int\limits_0^{\infty}x^{\nu-1}
e^{-\beta x^2}dx=
\frac {2^{-\nu} \beta^{-\frac {\nu} {2}}\sqrt{\pi}\;\Gamma(\nu)}
{\Gamma\left(\frac {\nu+1} {2}\right)},
\end{equation}
another important result that we will use in Section 3.


\setcounter{equation}{0}

\section{The $\nu$-dimensional BG distribution}

The BG partition function ${\cal Z}_\nu$ is
\begin{equation}
\label{ep3.1}
{\cal Z}_\nu=\int\limits_Me^{-\beta\left(\frac {p^2} {2m}-
\frac {GmM} {r}\right)}d^\nu xd^\nu p.
\end{equation}
For effecting the integration process one uses hyper-spherical coordinates and two integrals, each in  $\nu$ dimensions. The corresponding change of variables is defined as
\[x_1=r\cos\theta_1\]
\[x_2=r\sin\theta_1\cos\theta_2\]
\[x_3=r\sin\theta_1\sin\theta_2\cos\theta_3\]
\[\cdot\]
\[\cdot\]
\[x_{\nu-1}=r\sin\theta_1......\sin\theta_{\nu-2}\cos\theta_{\nu-1}\]
\begin{equation}
\label{ep3.2}
x_\nu=\sin\theta_1......\sin\theta_{\nu-1}\sin\theta_{\nu-1},
\end{equation} 
where $0\leq\theta_j\leq\pi$, $1\leq j\leq\nu-2$, and $0\leq\theta_{\nu-1}\leq 2\pi$.
The integration on the angular variables ($\Omega_\nu=(\theta_1, \theta_2,...,\theta_{\nu-1})$)
yields as a result
\begin{equation}
\label{ep3.3}
\int\limits_{\Omega_\nu}d\Omega_\nu=\left[\frac {2\pi^{\frac {\nu} {2}}} {\Gamma\left(
\frac {\nu} {2}\right)}\right]
\end{equation}
Ones is left then with just two radial coordinates (one in $r-$ space and the other in $p-$ space) and 
$2(\nu -1)$ angles. Accordingly, 
\begin{equation}
\label{ep3.4}
{\cal Z}_\nu=
\left[\frac {2\pi^{\frac {\nu} {2}}} {\Gamma\left(
\frac {\nu} {2}\right)}\right]^2
\iint\limits_0^{\infty}
(rp)^{\nu-1}
e^{-\beta\left(\frac {p^2} {2m}-
\frac {GmM} {r}\right)}dr\;dp.
\end{equation}
Now, using (\ref{ep2.9}) for
$\int\limits_0^{\infty}
r^{\nu-1}
e^{\beta\frac {GmM} {r}}dr$
and (\ref{ep2.13}) for
$\int\limits_0^{\infty}
p^{\nu-1}
e^{-\beta\frac {p^2} {2m}}dp$
we obtain
\begin{equation}
\label{ep3.5}
{\cal Z}_\nu=4\sqrt{\pi}\cos(\pi\nu)
\left(\frac {\pi^2\beta G^2m^3M^2} {2}\right)^{\frac {\nu}{2}}
\frac {\Gamma(\nu)\Gamma(-\nu)}
{\left[\Gamma\left(\frac {\nu} {2}\right)\right]^2
\Gamma\left(\frac {\nu+1} {2}\right)}.
\end{equation}
From (\ref{ep3.5}) one gathers that poles appear for any dimension $\nu$, 
 $\nu=3$ included. Thus, appeal to dimensional regularization (DR) will be mandatory. To this effect we will use in Section 4 the   DR-Bollini @ Giambiagi's technique's generalization given in \cite{tp1}. \vskip 3mm 
 
\nd Before, we still need an expression for  the mean energy 
\begin{equation}
\label{ep3.6}
<{\cal U}>_\nu=\frac {1} {{\cal Z}_\nu}
\int\limits_Me^{-\beta\left(\frac {p^2} {2m}-
\frac {GmM} {r}\right)}
\left(\frac {p^2} {2m}-\frac {GmM} {r}\right)d^\nu xd^\nu p.
\end{equation}
Appealing to the hyper-spherical coordinates previously mentioned we obtain for $<{\cal U}>_\nu$ 
\begin{equation}
\label{ep3.7}
<{\cal U}>_\nu=
\frac {1} {{\cal Z}_\nu}
\left[\frac {2\pi^{\frac {\nu} {2}}} {\Gamma\left(
\frac {\nu} {2}\right)}\right]^2
\iint\limits_o^{\infty}e^{-\beta\left(\frac {p^2} {2m}-
\frac {GmM} {r}\right)}
\left(\frac {p^2} {2m}-\frac {GmM} {r}\right)p^{\nu-1}r^{\nu-1} dp\;dr.
\end{equation}
At this stage we use again (\ref{ep2.9}) 
and (\ref{ep2.13}), which yields for the mean energy 
\[<U>_\nu=\frac {1} {{\cal Z}_\nu}
\frac {\sqrt{\pi}} {\beta}\cos(\pi\nu)
\left(\frac {\pi^2\beta G^2m^3M^2} {2}\right)^{\frac {\nu}{2}}\otimes\]
\begin{equation}
\label{ep3.8}
\left[
\frac {\Gamma(\nu+2)\Gamma(-\nu)}
{\left[\Gamma\left(\frac {\nu} {2}\right)\right]^2
\Gamma\left(\frac {\nu+3} {2}\right)}+4
\frac {\Gamma(\nu)\Gamma(1-\nu)}
{\left[\Gamma\left(\frac {\nu} {2}\right)\right]^2
\Gamma\left(\frac {\nu+1} {2}\right)}\right].
\end{equation}


\section{The 3D regularized BG distribution}

\setcounter{equation}{0}

We go back to (\ref{ep3.5}). The idea it to work out the ensuing dimensional regularization (DR) process.
If we have, for instance, an expression $F(\nu)$ that diverges, say, for $\nu=3$, 
our 
Bollini-Giambiagi's DR  generalized approach consists in performing the 
Laurent-expansion of $F$ around $\nu=3$ and select
afterwards, as the physical result for $F$, the $\nu=3$-independent term in the
expansion. The justification for such a procedure is clearly explained in \cite{tp1}.\vskip 3mm

\nd In our case, the corresponding Laurent expansion in the variable $\nu$ around $\nu=3$ is
\[{\cal Z}_\nu=-\frac {2} {3\sqrt{\pi}}
\frac {(2\pi^2\beta G^2m^3M^2)^{\frac {3} {2}}} {3(\nu-3)}-
\frac {1} {3\sqrt{\pi}}
(2\pi^2\beta G^2m^3M^2)^{\frac {3} {2}}\otimes\]
\begin{equation}
\label{ep5.1}
\left[
\ln\left(2\pi^2\beta G^2m^3M^2\right)-
\boldsymbol{C}-\frac {17} {3}
\right]+
\sum\limits_{s=1}^{\infty}a_s(\nu-3)^s.
\end{equation} 
where $\boldsymbol{C}$ is  Euler's constant. We clearly see that ${\cal Z}_\nu$
diverges at $\nu=3$.
By definition (and this is the essence of DR), the independent $(\nu-3)$-term
in the ${\cal Z}_\nu$-Laurent expansion yields the physical value of the ${\cal Z}$.
Thus, 
\begin{equation}
\label{ep5.2}
{\cal Z}=\frac {1} {3\sqrt{\pi}}(2\pi^2\beta G^2m^3M^2)^{\frac {3} {2}}\left[
\frac {17} {3}-\boldsymbol{C}-
\ln\left(8\pi^2\beta G^2m^3M^2\right)\right].
\end{equation}
Since ${\cal Z}$ must be positive, one faces a temperature-lower bound
\begin{equation}
\label{eq5.3}
T>\frac {e^{-\frac {17} {3}-\boldsymbol{C}}} {k_B}8\pi^2G^2m^3M^2.
\end{equation}
\nd 
Similarly,  from (\ref{ep3.8}), we  have for $<{\cal U}>$ the Laurent expansion
\[{\cal Z}<{\cal U}>_\nu=\frac {8} {\sqrt{\pi}\beta(\nu-3)}
\left(\frac {\pi^2\beta G^2m^3M^2} {2}\right)^{\frac {3} {2}} +
\frac {8} {\sqrt{\pi}\beta}
\left(\frac {\pi^2\beta G^2m^3M^2} {2}\right)^{\frac {3} {2}}\otimes\]
\begin{equation}
\label{ep5.4}
\left[\frac {1} {2}
\ln\left(\frac {\pi^2\beta G^2m^3M^2} {2}\right)+
2\ln 2-\frac {\boldsymbol{C}} {2} -\frac {5} {2}\right]+
\sum\limits_{s=1}^{\infty}a_s(\nu-3)^s.
\end{equation}
where ${\cal Z}$ is given by (\ref{ep5.2}).
Accordingly, the $(\nu-3)$-independent term is the physical value of $<{\cal U}>$
\begin{equation}
\label{ep5.5}
<{\cal U}>=\frac {1} {{\cal Z}}
\frac {8} {\sqrt{\pi}\beta}
\left(\frac {\pi^2\beta G^2m^3M^2} {2}\right)^{\frac {3} {2}}\left[
\frac {1} {2}
\ln\left(\frac {\pi^2\beta G^2m^3M^2} {2}\right)+
2\ln 2-\frac {\boldsymbol{C}} {2}-\frac {5} {2}\right],
\end{equation}
Replacing here the physical value of ${\cal Z}$ given by (\ref{ep5.2}) we  now  obtain
\begin{equation}
\label{ep5.6}
<{\cal U}>=-
\frac {3} {2\beta}
\frac {\ln\left(\pi^2\beta G^2m^3M^2\right)+
3\ln2-\boldsymbol{C}-5}
{\ln\left(\pi^2\beta G^2m^3M^2\right)+\ln 8-
\boldsymbol{C}-\frac {17} {3}}.
\end{equation}

\section{Specific Heat}

\setcounter{equation}{0}

We are now in possession, for the first time ever, of a canonical gravitational
mean energy function. Thus,  we use it for evaluating the specific heat 
 
${\cal C}=\frac {\partial<{\cal U}>} {\partial T}$. 
Thus, we obtain
\[{\cal C}=\frac {3k} {2}\frac {\ln(\pi^2\beta G^2m^3M^2)+3\ln 2
-6--\boldsymbol{C}}
{\frac {17} {3}+\boldsymbol{C}-\ln(2\pi^2\beta G^2m^3M^2)-\ln 2}-\]
\begin{equation}
\label{ep7.1}
\frac {3k} {2}
\frac {\ln(16\pi^2\beta G^2m^3M^2)+3\ln 2-5-\boldsymbol{C}}
{\left[\frac {17} {3}+\boldsymbol{C}-
\ln(2\pi^2\beta G^2m^3M^2)-\ln 2\right]^2}
\end{equation}

\nd Figs. 1 depict the specific heat corresponding
to Eq.  (\ref{ep7.1}). We call 
$E=G^2m^3M^2$ with $m<<<M$. We express quantities in $k_BT/E$-units.  The specific heat is negative, as befits gravitation \cite{lb}. 
\nd Indeed, such an
occurrence has been associated to self-gravitational systems
\cite{lb}. Thirring has magnificently illustrated on negative heat capacities \cite{thirring}. In turn, Verlinde has associated this type of
systems to an entropic force \cite{verlinde}. It is natural to
conjecture then that such a force may appear at the energy-associated poles. 
 Notice also that temperature ranges are restricted. There is
a $T-$lower bound.

\begin{figure}[h]
\begin{center}
\includegraphics[scale=0.6,angle=0]{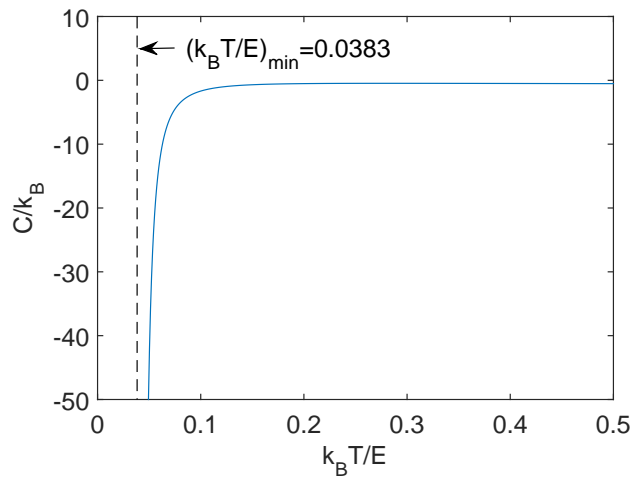}
\vspace{-0.2cm} \caption{Specific heat versus 
$k_BT/E$}\label{fig1}
\end{center}
\end{figure}

\section{Discussion}

\setcounter{equation}{0}

\nd  It is commonly believed that the partition function
${\cal Z}$ associated to a Boltzmann-Gibbs (BG) probability distribution diverges \cite{lb,grav1}.
\vskip 3mm
\nd However, such belief does not take into account the
possibility of  analytical extensions. We have conclusively shown here  that analytical extension coupled
to dimensional regularization (DR), allows one to obtain a finite gravitational
BG partition function. 

\vskip 3mm \nd We acknowledge the fact that the classical gravitational problem 
has wider horizons, that were not  touched here. Our contribution was just that of providing a finite partition function for the two-body gravitational problem.

\vskip 3mm\nd A special point to be remarked is the following.
The statistical gravitational problem is one in which 
the BG treatment is considerably more involved than
its Tsallis counterpart. The latter needs only dimensional regularization,
the former requires, in addition, analytical extension.

\vskip 3mm\nd Note that dealing with Newton's gravity with Tsallis's q-statistics plus
the DR also solves the problem of obtaining a
for $q=4/3$ \cite{z} . To do the same with BG-statistics demands, in addition, analytical extension. One may 
wonder what is the role played by the parameter $q$. We have shown in the
references given in \cite{q1} that $q$ is an indicator of the
energy-amount  involved in physical processes related to 
resonances and  Quantum Field Theory (QFT). The greater is the q-value,
the larger the value of the energy involved in the process. According to
results of   
the Alice experiment of the LHC \cite{q1}, one finds that  non-linear
quantum fields would manifest themselves around 15 TEVs and that these
fields would eventually correspond to an approximate value of q = 1.5. The
value $q = 1$ would correspond the usual, linear QFT.

\vskip 3mm\nd One might perhaps conjecture that for Newton's  gravity (NG) something similar happens. 
For usual energies, the NG-statistical treatment should be the BG one.
At bigger energies, one may better resort  to Tsallis statistics. A relevant example is given
in Ref. \cite{aparp}.

\end{document}